\documentclass[aps, nofootinbib, preprint, showkeys, floatfix]{revtex4} 
\usepackage[pdftex]{graphicx}
\usepackage{amsmath, amssymb, latexsym}
\usepackage{slashed}
\usepackage{url}
\usepackage{multirow}
\def\vev#1{\left\langle #1\right\rangle}
\catcode`\@=11
\def\lsim{\mathrel{\mathpalette\@versim<}}
\def\gsim{\mathrel{\mathpalette\@versim>}}
\def\@versim#1#2{\vcenter{\offinterlineskip
\ialign{$\m@th#1\hfil##\hfil$\crcr#2\crcr\sim\crcr } }}
\catcode`\@=12

\newcommand{\AddrUNAM}{ {\it Instituto de F\'{\i}sica, Universidad Nacional Aut\'onoma de M\'exico, A.P. 20-364, Ciudad de M\'exico 01000, M\'exico.}}

\begin{document}
\title{Seesaw scale discrete dark matter and two-zero texture Majorana neutrino mass matrices}
\author{J. M. Lamprea} \email{jmlamprea@fisica.unam.mx}\affiliation{\AddrUNAM}
\author{E. Peinado} \email{epeinado@fisica.unam.mx}\affiliation{\AddrUNAM}
%\pacs{14.60.Pq, 12.60.Fr, 14.80.-j}
\begin{abstract}
In this paper we present a scenario where the stability of dark matter and the phenomenology of neutrinos are related by the spontaneous breaking of a non-Abelian flavor symmetry ($A_4$). In this scenario the breaking is done at the seesaw scale, in such a way that  what remains of the flavor symmetry is a $Z_2$ symmetry, which stabilizes the dark matter. We have proposed two models based on this idea, for which we have calculated their neutrino mass matrices achieving two-zero texture in both cases. Accordingly, we have updated this two-zero texture phenomenology finding an interesting correlation between the reactor mixing angle and the sum of the light neutrino masses. We also have a correlation between the  lightest neutrino mass and the neutrinoless double beta decay effective mass, obtaining a lower bound for the effective mass within the region of the nearly future experimental sensitivities. 
\end{abstract}
\maketitle
\section{Introduction}
Neutrino masses, the existence of dark matter (DM), and the baryon asymmetry in the Universe (BAU) are the most important evidences of physics beyond the Standard Model (SM). Here, we propose that the same symmetry explaining neutrino mixing angles is also responsible for the dark matter stability in the context of the discrete dark matter (DDM) mechanism~\cite{Hirsch:2010ru}. Under certain conditions it would be possible also to account for the BAU via leptogenesis. The DDM is based upon the fact that the breaking of a discrete non-Abelian flavor symmetry into one of its subgroups by means of the electroweak symmetry breaking mechanism\footnote{It is possible to break the flavor symmetry with flavon field at other scale other than the electroweak, as we will discuss later.}, accounts for the neutrino masses and mixing pattern and for the dark matter stability. 

In the original model~\cite{Hirsch:2010ru},  the group of even permutation of four objects, $A_4$, was considered as the flavor symmetry.\footnote{ The motivation for choosing $A_4$ as the flavor group is because it is the smallest non-Abelian discrete group with triplet irreducible representation. Therefore, it is possible to have in the same multiplet some inert and active particles after the flavor symmetry breaking and at the same time a reduced number of couplings for  these triplets.  Later we will see that this reduced number of couplings is the reason why we got correlations between the observables in the neutrino sector.} $A_4$ contains one tridimensional irreducible representation, ${\bf 3}$,  and three one-dimensional irreducible representations,  $\bf{1}$,  $\bf{1^\prime}$, and $\bf{1^{\prime\prime}}$,  whose algebra can be reviewed in the Appendix, see, for instance,~\cite{Ma:2001dn,Babu:2002dz,Altarelli:2005yx,Altarelli:2006kg}.  The particle content includes four $SU(2)$  Higgs doublets, three of them transforming as  an $A_4$ triplet $\eta = (\eta_1,\, \eta_2,\, \eta_3)$ and the SM Higgs $H$ as the singlet  $\bf{1}$; four right-handed (RH) neutrinos, three of them in a triplet representation of $A_4$, $N_T = (N_1, N_2, N_3)$; and  a singlet ${\bf 1}$ of $A_4$, $N_4$. The lepton doublets $L_i$ and the right charged leptons $l_i$ transform as the three different singlets under  $A_4$, in such a way that the mass matrix for the charged lepton is diagonal.  Breaking the $A_4$ symmetry into a $Z_2$ subgroup through the electroweak symmetry breaking, provides the stability mechanism for the DM, arising from the $Z_2$-odd  part of the triplet $\eta$  and at the same time accounts for the neutrino masses and mixing patterns by means of the type I seesaw~\cite{Minkowski:1977sc,Yanagida:1979as,Glashow,GellMann:1980vs,Mohapatra:1980yp,schechter:1980gr}. The predictions for the neutrino sector are an inverse mass hierarchy spectrum with a massless neutrino, $m_{\nu_3} = 0$\footnote{This is because  only  two of the  RH neutrinos participate in the seesaw mechanism.}, and a vanishing reactor neutrino mixing angle, $\theta_{13} = 0$, that nowadays is ruled out by the current experimental data~\cite{Tang:2015vug,Forero:2014bxa,Gonzalez-Garcia:2015qrr,Capozzi:2016rtj}. Even though, there is an issue with the reactor mixing angle, the original $A_4$ model has a quite interesting neutrino and DM phenomenology, as can be seen in~\cite{Boucenna:2011tj}.

In a subsequent paper, this $A_4$ model has been modified~\cite{Meloni:2010sk}, by adding a fifth RH neutrino $N_5$ transforming as $\bf{1^{\prime\prime}}$ and changing the representation of $N_4$ to $\bf{1^{\prime}}$. This new model gives as predictions: a normal mass spectrum, a lower bound for the neutrinoless double beta decay effective mass, $|m_{ee}|$, and a nonzero reactor neutrino mixing angle. Nevertheless, even if this mixing angle were nonzero at its maximum value, it is again ruled out by the current experimental data~\cite{Tang:2015vug}. 

There are some other works in this direction, where other flavor symmetry groups have been used, for instance, a model based on the dihedral group $D_4$ where some flavor changing neutral currents are present and constrain the DM sector~\cite{Meloni:2011cc} and a model based on the $\Delta(54)$ ~\cite{Boucenna:2012qb}. Finally,  it is worth mentioning that there have been works tackling the problem of the vanishing reactor mixing angle within the $A_4$ DDM model~\cite{Hamada:2014xha}, but in such a case the $A_4$ symmetry has to be explicitly broken in the scalar potential.  For models in which dark matter transforms nontrivially under a non-Abelian flavor symmetry, see, for instance,~\cite{Adulpravitchai:2011ei, Varzielas:2015joa,Varzielas:2015sno}.

The paper is organized as follows: In Sec.~\ref{sec:Models} we explain our models giving their matter content and derive the neutrino mass matrices. In Sec.~\ref{sec:ModRes} we discuss the phenomenology, and in Sec.~\ref{sec:Con} we draw our conclusions.  

\section{Reactor mixing angle and the DDM mechanism}
\label{sec:Models}
We will consider two extensions of the model in Ref.~\cite{Hirsch:2010ru}, hereafter referred as model A and model B, where in addition to the original model matter content, we have added one extra RH neutrino  $N_5$, in a singlet representation  of $A_4$ ($\bf{1^{\prime}}$ or $\bf{1^{\prime\prime}}$), and three real scalar singlets of  the SM  transforming as a triplet under $A_4$, $\phi= (\phi_1,~\phi_2,~ \phi_3) $. The relevant particle content and quantum numbers of model A and model B are summarized in Tables~\ref{tab:ModA} and \ref{tab:ModB}, respectively. The $N_5$ RH neutrino  is assigned to the  $\bf{1^{\prime}}$ representation of $A_4$ in model A and to the $\bf{1^{\prime\prime}}$ $A_4$ representation in model B. The flavon fields, $\phi $, acquire a vacuum expectation value around the seesaw scale, such that  $A_4$ is broken into a $Z_2$ at this scale instead of at the electroweak scale as in the original model. In this way, the flavon fields contribute to the RH neutrino masses.

\subsection*{Model A}
\begin{table}[h!]
\begin{center}
\begin{tabular}{|c|c|c|c|c|c|c|c|c|c||c|c|c|}
\hline
& $\,L_e\,$ & $\,L_\mu \,$ & $\,L_\tau \,$ & $\,\,l_e^c\,\,$ & $\,\,l_\mu^c\,\,$ & $\,\,l_\tau^c\,\,$ & $N_T\,$ & $\,N_4\,$ & $\,N_5\,$ & $\,H\,$ & $\,\eta \,$ & $\,\phi \,$\\
\hline
SU(2) & 2 & 2 & 2 & 1 & 1 & 1 & 1 & 1 & 1 & 2 & 2 & 1\\
\hline
$A_4$ & 1 & $1^\prime$ & $1^{\prime\prime}$ & 1 & $1^{\prime\prime}$ & $1^\prime$ & 3 & 1 & $1^\prime$ & 1 & 3  &  3 \\
\hline
\end{tabular}\caption{\it Summary of the relevant particle content and quantum numbers for model A.} \label{tab:ModA}
\end{center}
\end{table}
If we consider the matter content in Table ~\ref{tab:ModA},  the lepton Yukawa Lagrangian is given by\footnote{The contribution $y_1^N [N_T \,\phi]_3 N_T$ accounts for the symmetric part of how the two triplets can be contracted, namely $[N_T \,\phi]_{3_1}$ and $[N_T \,\phi]_{3_2}$.}
\begin{align}
\label{eq:YukA}
\mathcal{L}^{(\text{A})}_{\text{Y}} &= 
y_e L_e l_e^c H + y_\mu L_\mu l_\mu^c H + y_\tau L_\tau l_\tau^c H\nonumber\\
&+ y_1^\nu L_e [N_T\, \eta]_1 + y_2^\nu L_\mu [N_T\, \eta] _{1''} + y_3^\nu L_\tau [N_T\, \eta]_{1'} + y_4^\nu L_e \,N_4\, H + y_5^\nu L_\tau\, N_5\, H\\
&+ M_1\, N_T N_T + M_2\, N_4 N_4 + y_1^N [N_T \,\phi]_3 N_T +  y_2^N [N_T\, \phi]_{1} N_4 + y_3^N [N_T\, \phi]_{1''} N_5 + \textit{H.c.}\nonumber
\end{align}
where $[a,\,b]_{j},$ stands for the product of the two triplets $a, \,b$ contracted into the $j$  representation of $A_4$. In this way, $H$ is responsible for quark (considering the quarks as singlet of $A_4$) and charged lepton masses, the latter automatically diagonal,  $M_l = v_h\,\text{diag} \left(\,y_e,\, y_\mu,\, y_\tau\, \right)$. The Dirac neutrino mass matrix arises from $H$ and $\eta$. The flavon fields will contribute to the RH neutrino mass matrix. Once the flavon fields acquire a vev, $A_4$ will be broken. In order to preserve a $Z_2$ symmetry, the alignment of the vevs will be of the form
\begin{equation}
\label{eq:Vevs}
\vev{H^0} = v_h \ne 0,\qquad
\vev{ \eta^0_1} = v_\eta \ne 0,\qquad
\vev{\eta^0_{2,3}} = 0,  \qquad
\vev{ \phi_1} = v_\phi \ne 0,\qquad
\vev{\phi_{2,3}} = 0.
\end{equation} 
Therefore, $(1,\,0,\,0)$ is the vacuum alignment for the $A_4$ scalar triplets, which is a way to break spontaneously $A_4$ into a $Z_2$ subgroup, in the $A_4$ basis where the $S$ generator is diagonal, see the Appendix.

From Eqs. (\ref{eq:YukA}) and~(\ref{eq:Vevs}), the Dirac neutrino mass matrix is given by
\begin{equation}
\label{eq:MDirA}
 m_{\text{D}}^{(\text{A})} = 
  \begin{pmatrix}
  y_1^\nu  v_\eta & 0 & 0 & y_4^\nu  v_h & 0\\
  y_2^\nu  v_\eta & 0 & 0 & 0 & 0\\
  y_3^\nu  v_\eta & 0 & 0 & 0 & y_5^\nu v_h
  \end{pmatrix},
\end{equation}
and the Majorana neutrino mass matrix is
\begin{equation}
\label{eq:MassR}
  M_{\text{R}} = 
  \begin{pmatrix}
  M_1 & 0 & 0 & y_2^N v_\phi & y_3^N v_\phi\\
  0 & M_1 & y_1^N v_\phi & 0 & 0\\
  0 & y_1^N v_\phi & M_1 & 0 & 0\\
  y_2^N v_\phi & 0 & 0 & M_2 & 0 \\
  y_3^N v_\phi & 0 & 0 & 0 & 0 
  \end{pmatrix}.
\end{equation}
With these mass matrices, the light neutrinos get Majorana masses through the type I seesaw relation,  $m_\nu = - m_{\text{D}_{3\times 5}} M_{\text{R}_{5\times 5}}^{-1} m_{\text{D}_{3\times 5}}^\text{T}$, taking the form
\begin{equation} 
\label{eq:MnuA}
m_\nu^{(\text{A})}
\equiv
\begin{pmatrix}
a & 0 & b\\
0 & 0 & c\\
b & c & d
\end{pmatrix},
\end{equation}
where
\begin{equation}
\label{eq:ab}
\begin{array}{ll}
a = \frac{(y_4^\nu v_h)^2 }{M_2}, &
b = \frac{y_1^\nu y_5^\nu v_\eta v_h }{y_3^N v_\phi} - \frac{ y_2^N y_4^\nu y_5^\nu v_h^2 }{y_3^N M_2 }, \\ \\
c = \frac{y_2^\nu y_5^\nu v_\eta v_h}{y_3^N v_\phi}, &
d = \frac{ (y_2^N y_5^\nu v_h)^2 }{(y_3^N)^2 M_2 } - \frac{ (y_5^\nu v_h)^2  M_1}{(y_3^N v_\phi)^2} + 2\ \frac{y_3^\nu y_5^\nu v_\eta v_h}{y_3^N v_\phi}.
\end{array}\end{equation} 
The mass matrix in Eq. (\ref{eq:MnuA}) has the form of the $B_3$ two-zero neutrino mass matrix~\cite{Frampton:2002yf}, which phenomenology has been extensively studied in the literature, see, for instance,~\cite{Frampton:2002yf,Xing:2002ta,Frampton:2002rn,Kageyama:2002zw,Merle:2006du, Fritzsch:2011qv, Ludl:2011vv,Meloni:2014yea,Zhou:2015qua,Kitabayashi:2015jdj}. This matrix is consistent with both neutrino mass hierarchies and can accommodate the experimental value for the reactor mixing angle, $\theta_{13}$~\cite{Frampton:2002yf,Xing:2002ta,Frampton:2002rn,Kageyama:2002zw,Merle:2006du, Fritzsch:2011qv, Ludl:2011vv,Meloni:2014yea,Zhou:2015qua,Kitabayashi:2015jdj}. The phenomenological implications of this scenario are studied in Sec.~\ref{sec:ModRes}.

\subsection*{Model B}
\begin{table}[h!]
\begin{center}
\begin{tabular}{|c|c|c|c|c|c|c|c|c|c||c|c|c|}
\hline
& $\,L_e\,$&$\,L_{\mu}\,$&$\,L_{\tau}\,$&$\,\,l_{e}^c\,\,$&$\,\,l_{{\mu}}^c\,\,$&$\,\,l_{{\tau}}^c\,\,$&$N_{T}\,$&$\,N_4\,$&$\,N_5\,$&$\,H\,$&$\,\eta\,$&\,$\phi\,$\\
\hline
SU(2) & 2 & 2 & 2 & 1 & 1 & 1 & 1 & 1 & 1 & 2 & 2 & 1\\
\hline
$A_4$ & 1 & $1^\prime$ & $1^{\prime\prime}$ & 1 & $1^{\prime\prime}$ & $1^\prime$ & 3 & 1 & $1^{\prime\prime}$ & 1 & 3 & 3\\
\hline
\end{tabular}
\caption{\it Summary of the relevant particle content and quantum numbers for model B.}\label{tab:ModB}
\end{center}
\end{table}
The lepton Yukawa Lagrangian for the matter content and assignments of model B, in Table~\ref{tab:ModB}, is given by
\begin{align}
\label{eq:YukB}
\mathcal{L}^{(\text{B})}_{\text{Y}} &= 
y_e L_e l_e^c H + y_\mu L_\mu l_\mu^c H + y_\tau L_\tau l_\tau^c H \nonumber \\
 &+ y_1^\nu L_e [N_T\, \eta]_1 + y_2^\nu L_\mu [N_T\, \eta] _{1''} + y_3^\nu L_\tau [N_T\, \eta]_{1'} + y_4^\nu L_e \,N_4\, H + y_5^\nu L_\mu\, N_5\, H\\
 &+ M_1\, N_T N_T + M_2\, N_4 N_4 + y_1^N [N_T \,\phi]_{3} N_T + y_2^N [N_T\, \phi]_{1} N_4 + y_3^N [N_T\, \phi]_{1'} N_5 + \textit{H.c.}\nonumber
\end{align} 
As in model A, the mass matrix of the charged leptons is diagonal, due to the flavor symmetry, while the Dirac neutrino mass matrix takes the form
\begin{eqnarray}
\label{eq:MDirB}
 m_{\text{D}}^{(\text{B})} =
  \begin{pmatrix}
  y_1^\nu  v_\eta & 0 & 0 & y_4^\nu  v_h & 0\\
  y_2^\nu  v_\eta & 0 & 0 & 0 & y_5^\nu v_h\\
  y_3^\nu  v_\eta & 0 & 0 & 0 & 0
  \end{pmatrix}.
\end{eqnarray}
The Majorana neutrino mass matrix is of the same form as Eq.~(\ref{eq:MassR}).  
The light neutrinos mass matrix after the type I seesaw is
\begin{equation}
 \label{eq:MnuB}
m_\nu^{(\text{B})}
\equiv
\begin{pmatrix}
a & b & 0\\
b & d & c\\
0 & c & 0
\end{pmatrix},
\end{equation}
where 
\begin{equation}
\label{eq:cd}\begin{array}{ll}
a = \frac{(y_4^\nu v_h)^2 }{M_2}, &
b = \frac{y_1^\nu y_5^\nu v_\eta v_h }{y_3^N v_\phi} - \frac{ y_2^N y_4^\nu y_5^\nu v_h^2 }{y_3^N M_2 }, \\ \\
c = \frac{y_3^\nu y_5^\nu v_\eta v_h}{y_3^N v_\phi}, &
d = \frac{ (y_2^N y_5^\nu v_h)^2 }{(y_3^N)^2 M_2 } - \frac{ (y_5^\nu v_h)^2  M_1}{(y_3^N v_\phi)^2} + 2\ \frac{ y_2^\nu y_5^\nu v_\eta v_h}{ y_3^N v_\phi },
\end{array}
\end{equation} 
which correspond, as before, to another two-zero texture flavor neutrino mass matrix, $B_4$~\cite{Frampton:2002yf}, which also is consistent with both neutrino mass hierarchies and can also accommodate the reactor mixing angle, $\theta_{13}$~\cite{Frampton:2002yf,Xing:2002ta,Frampton:2002rn,Kageyama:2002zw,Merle:2006du,Fritzsch:2011qv, Ludl:2011vv,Meloni:2014yea,Zhou:2015qua,Kitabayashi:2015jdj}.

\section{Results}
\label{sec:ModRes}
In the previous section, we obtained the two-zero texture neutrino mass matrices $B_3$ and $B_4$ for models A and B, respectively. We performed the analysis using four independent constraints, coming from the two complex zeroes, to correlate two of the neutrino mixing parameters: the neutrino masses and mixing angles, the two Majorana phases and the Dirac CP violating phase.\footnote{The method we have used is known and can be reviewed, for instance, in~\cite{Ludl:2011vv,Meloni:2014yea}.} We took the experimental values of the three mixing angles and the two squared mass differences as inputs and numerically scanned within their $3\sigma$ regions and determined the regions allowed by two correlated variables of interest.  We have used in the analysis the data from three different groups that perform the neutrino global fits~\cite{Forero:2014bxa,Gonzalez-Garcia:2015qrr,Capozzi:2016rtj}.

\begin{figure}[!h]
\centering
\includegraphics[scale=0.37]{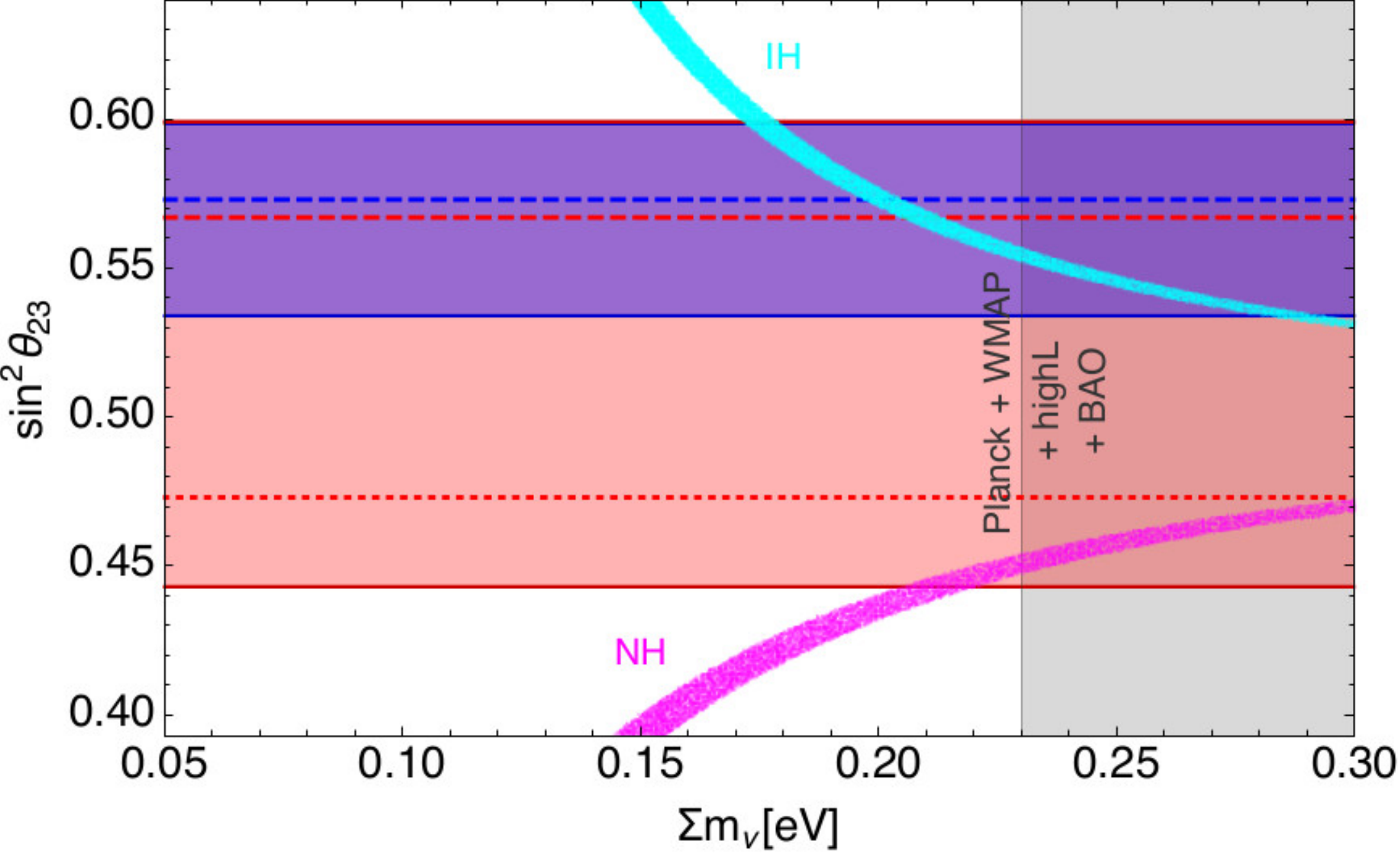} \includegraphics[scale=0.37]{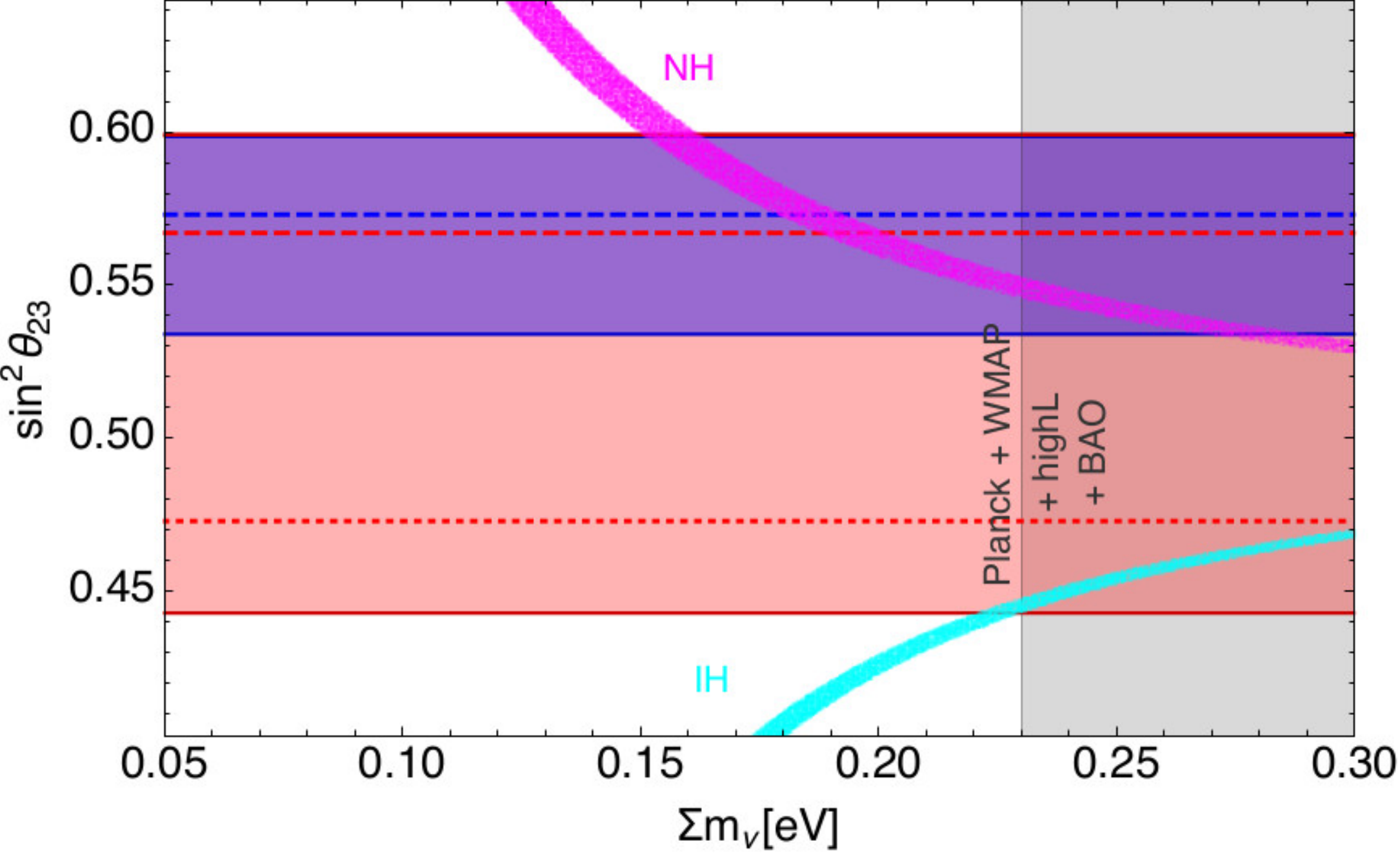}
\caption{Correlation between $\sin^2 \theta_{23}$ and the sum of the light neutrino masses, $\sum m_\nu$, in model A ($B_3$) on the left and model B ($B_4$) on the right, for NH (IH) in magenta (cyan). The horizontal red (blue) shaded region is the $1\sigma$ in $\sin^2 \theta_{23}$ for NH (IH). The red (blue) horizontal dashed line is the best fit value in NH (IH), and the doted red line is the value of local minimum in NH appearing in the data analysis used. The data was taken from~\cite{Forero:2014bxa}. The vertical grey shaded region is disfavored by the current Planck data~\cite{Ade:2015xua}.}
\label{fig:graf}
\end{figure}

\begin{figure}[!h]
\centering
\includegraphics[scale=0.37]{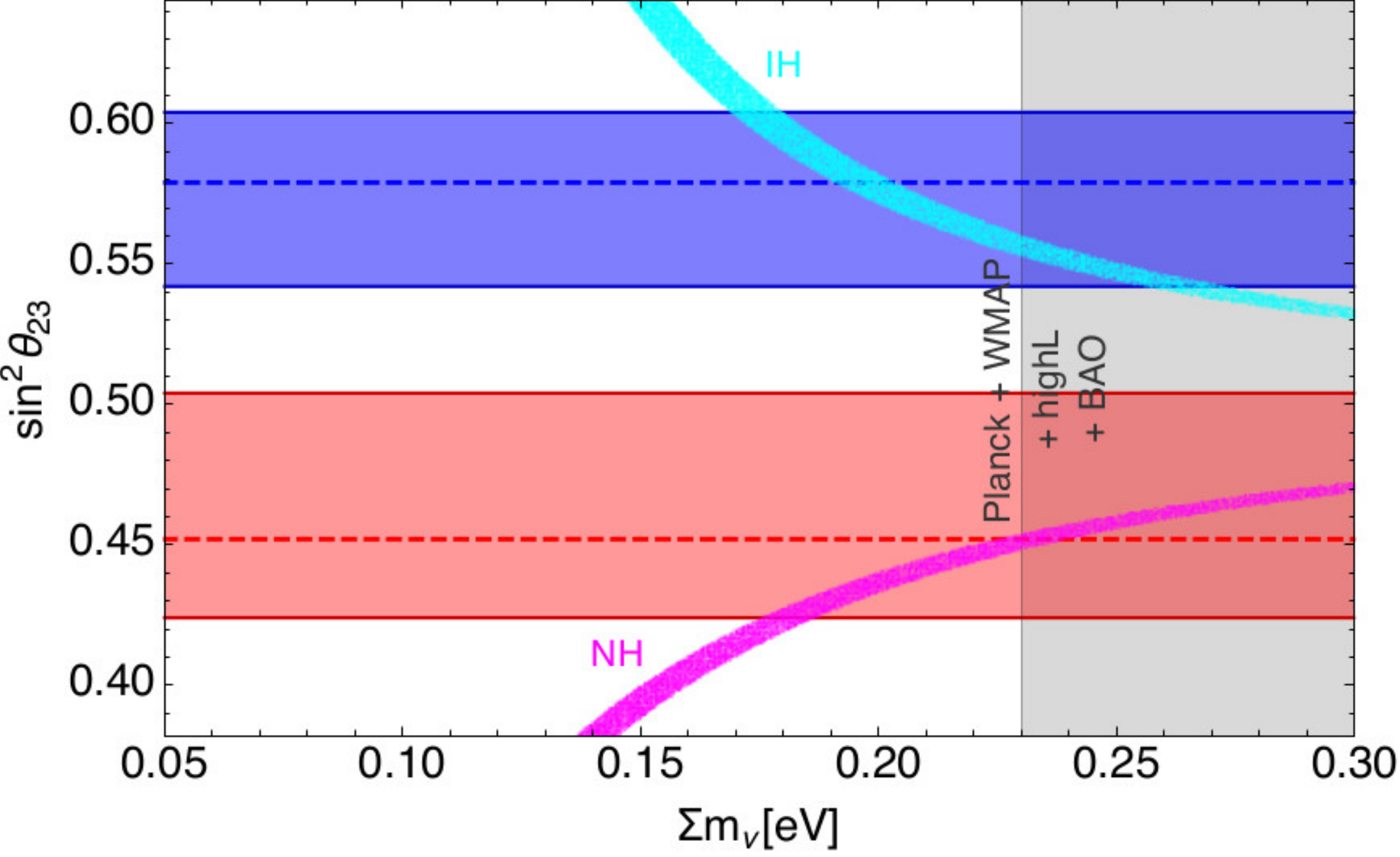} \includegraphics[scale=0.37]{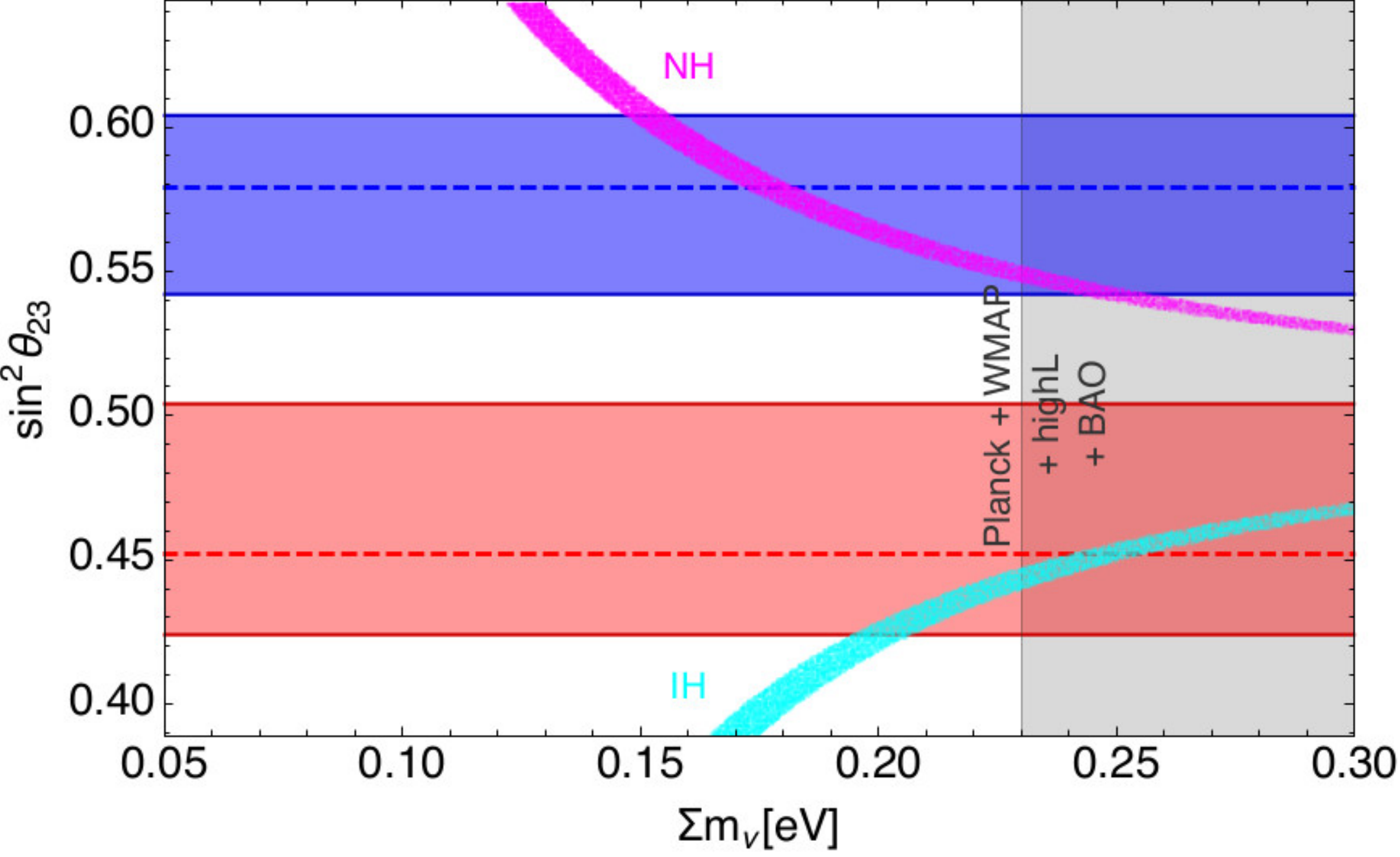}
\caption{Correlation between $\sin^2 \theta_{23}$ and the sum of the light neutrino masses, $\sum m_\nu$, in model A ($B_3$) on the left and model B ($B_4$) on the right, for NH (IH) in magenta (cyan). The red (blue) horizontal dashed line is the best fit value in NH (IH). The data was taken from~\cite{Gonzalez-Garcia:2015qrr}. The vertical grey shaded region is disfavored by the current Planck data~\cite{Ade:2015xua}.}
\label{fig:graf2}
\end{figure}

\begin{figure}[!h]
\centering
\includegraphics[scale=0.37]{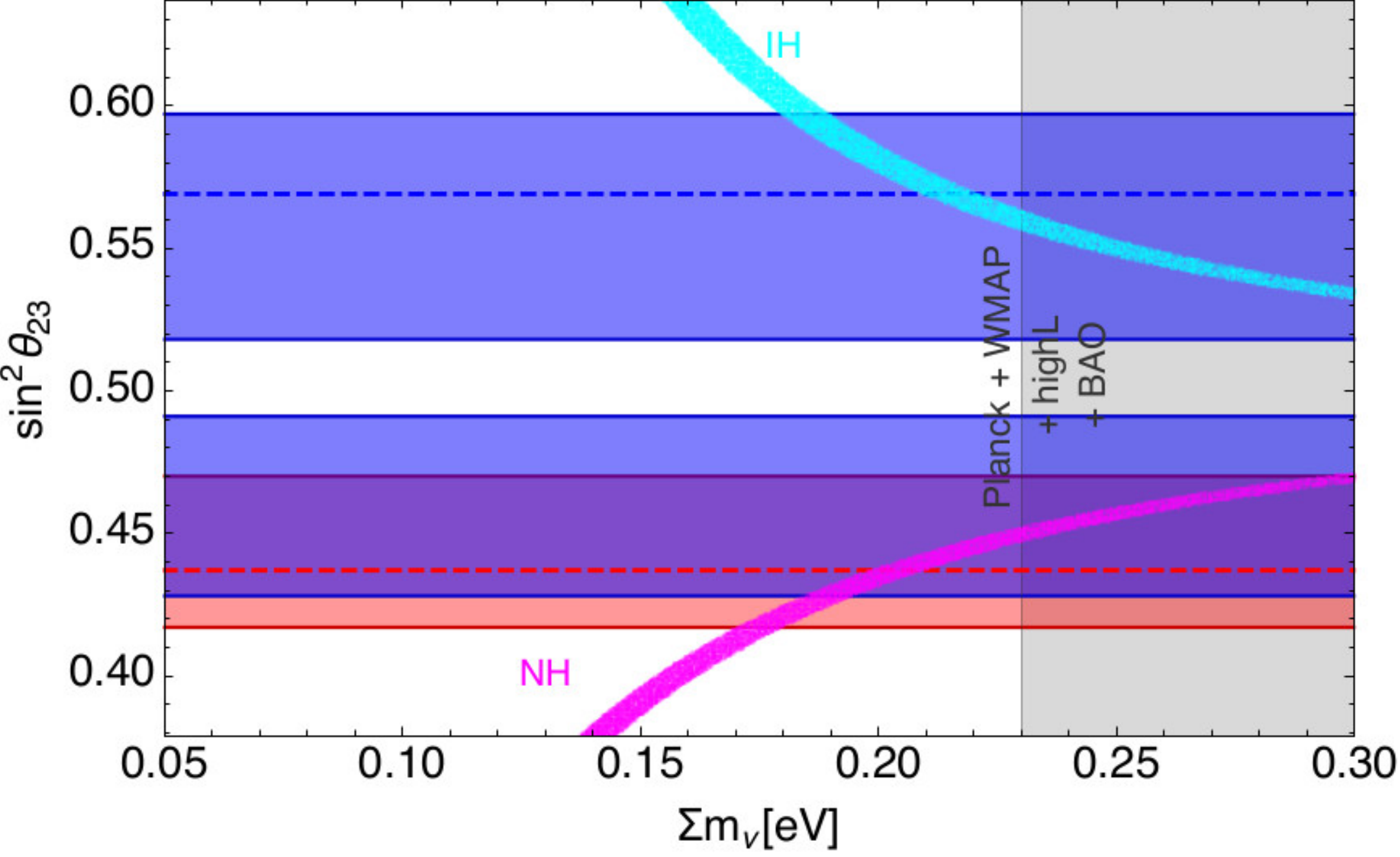} \includegraphics[scale=0.37]{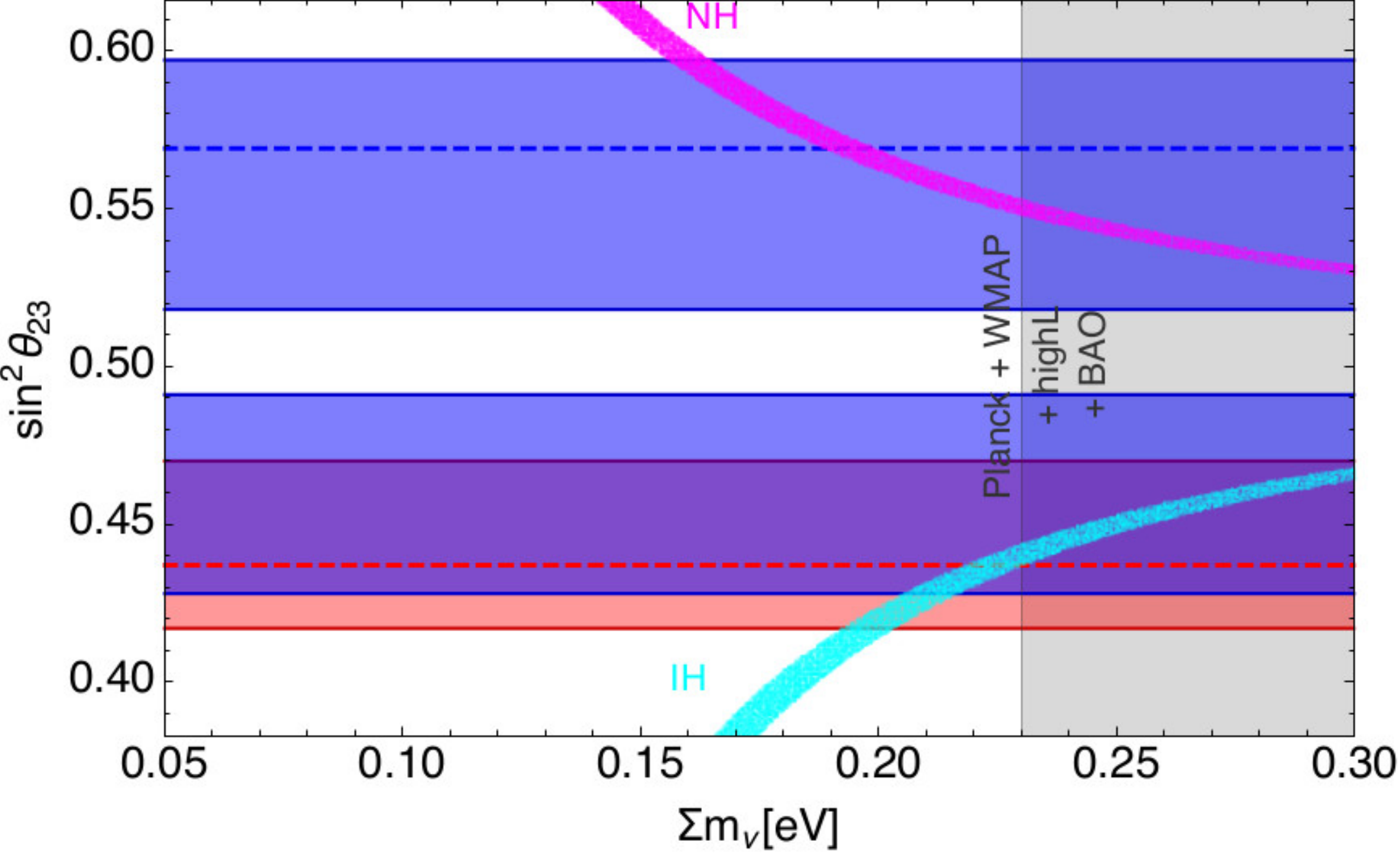}
\caption{Correlation between $\sin^2 \theta_{23}$ and the sum of the light neutrino masses, $\sum m_\nu$, in model A ($B_3$) on the left and model B ($B_4$) on the right, for NH (IH) in magenta (cyan). The horizontal red (blue) shaded regions are the $1\sigma$ in $\sin^2 \theta_{23}$ for NH (IH). The case for IH has two favored $1sigma$ regions according to the data used. The red (blue) horizontal dashed line is the best fit value in NH (IH). The data was taken from~\cite{Capozzi:2016rtj}. The vertical grey shaded region is disfavored by the current Planck data~\cite{Ade:2015xua}.}
\label{fig:graf4}
\end{figure}

In Figs.~\ref{fig:graf},~\ref{fig:graf2}, and~\ref{fig:graf4} we show the correlation between the atmospheric mixing angle, $\sin^2 \theta_{23}$, and the sum of light neutrino masses, $\textstyle \sum m_\nu=m_{\nu_{1}}+m_{\nu_{2}}+m_{\nu_{3}}$, for model A on the left panels and model B on the right ones. In these graphics, the allowed $3\sigma$ regions in $\sin^2 \theta_{23}$ vs. $\textstyle \sum m_\nu$  for the normal hierarchy (NH) is plotted in magenta and for the inverse hierarchy (IH) in cyan. The  $1\sigma$ in the atmospheric angle are represented by the horizontal blue and red shaded regions for the inverted and normal mass hierarchy, respectively, and the best fit values correspond to the horizontal blue and red dashed lines for the inverse and normal hierarchies respectively. In Forero {\it et al.}~\cite{Forero:2014bxa}, they have a local minimum in the atmospheric mixing angle for the IH analysis; that we represent as a red pointed line in Fig. \ref{fig:graf}. In addition, in the analysis by Capozzi  {\it et al.}~\cite{Capozzi:2016rtj}, they have obtained two different and separated $1\sigma$ regions in the atmospheric angle also for the IH; that we show as the double blue shaded horizontal bands in Fig.~\ref{fig:graf4}.  The grey vertical band represents a disfavored region in the sum of light neutrino masses, $\textstyle \sum m_\nu < 0.23$ eV, by the Planck Collaboration~\cite{Ade:2015xua}.
 
From the plots in Figs. \ref{fig:graf}, \ref{fig:graf2}, and \ref{fig:graf4}, it can be seen that in model A both hierarchies have an overlap  within the 1$\sigma$ region for the atmospheric mixing angle, while in model B depending on what data is used  the overlap not always exists.  For data from Forero {\it et al.}~\cite{Forero:2014bxa} and  Gonzalez-Garcia {\it et al.}~\cite{Gonzalez-Garcia:2015qrr}, only the NH the atmospheric mixing angle overlaps with the 1$\sigma$ region (even though for data from \cite{Gonzalez-Garcia:2015qrr} it happens for large neutrino masses disfavored by Planck). For data from Capozzi {\it et al.}~\cite{Capozzi:2016rtj}, only  in the IH case there is the 1$\sigma$ overlap in the second octant for the atmospheric mixing angle. Finally, it is worth mentioning that the NH and IH regions in model A are  the same but interchanged in model B. 

\begin{figure}[!h]
\centering
\includegraphics[scale=0.37]{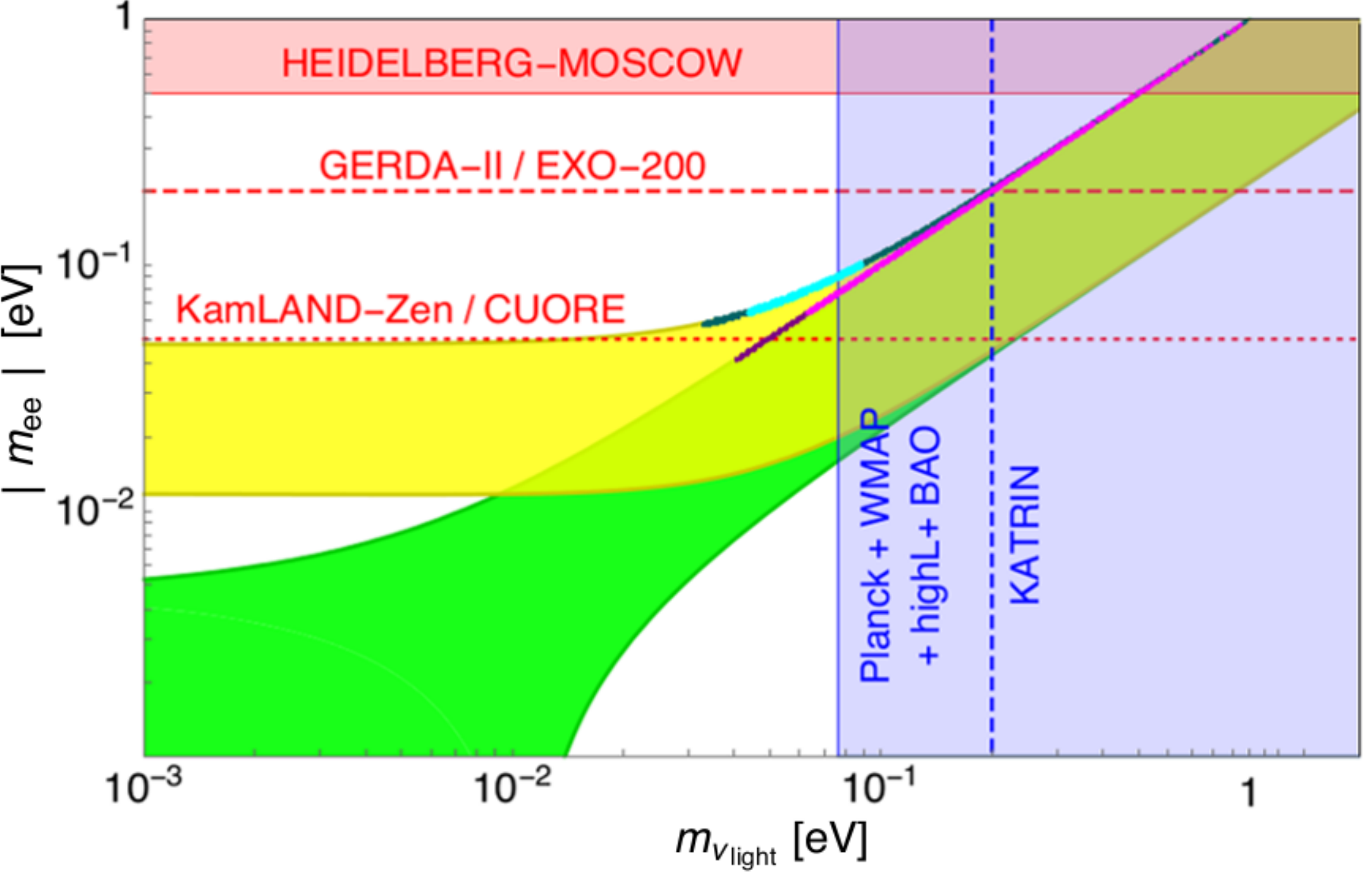} \includegraphics[scale=0.37]{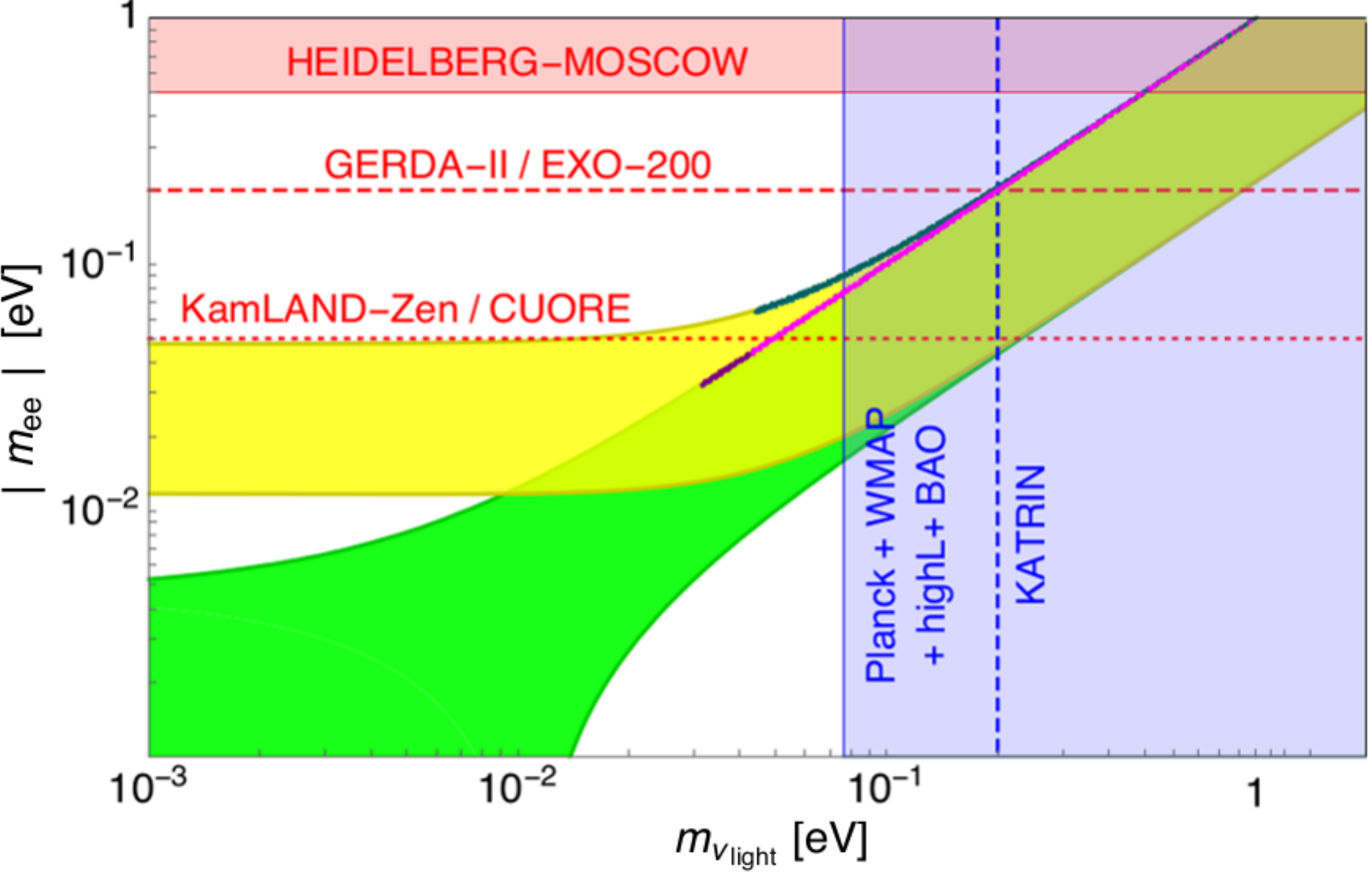}
\caption{Effective $0\nu\beta\beta$ parameter $|m_{ee}|$ versus the lightest neutrino mass $m_{\nu_\text{light}}$ in model A (B) on the left (right). In both models $m_{\nu_\text{light}}$ is $m_1$($m_3$) for NH (IH). The model allowed regions for NH are in magenta (dark magenta) for the $1\sigma$ ($3\sigma$)  atmospheric mixing angle region and for IH in cyan (dark cyan) for the $1\sigma$ ($3\sigma$)  on the atmospheric mixing angle region from~\cite{Forero:2014bxa}. The yellow (green) band correspond to the ``flavor-generic" inverse (normal) hierarchy neutrino spectra for $3\sigma$. The horizontal red shaded region is the current experimental limit on $0\nu\beta\beta$,  and the red (blue) horizontal (vertical) lines are the forthcoming experimental sensitivity on $|m_{ee}|$ ($m_{\nu_\text{light}}$), see~\cite{Agostini:2013mzu,Albert:2014awa,Gando:2012zm,Bornschein:2003xi,CUORE,Guiseppe:2008aa}.  The vertical blue shaded regions are disfavored by the current Planck data~\cite{Ade:2015xua}.}
\label{fig:graf1}
\end{figure}

\begin{figure}[!h]
\centering
\includegraphics[scale=0.37]{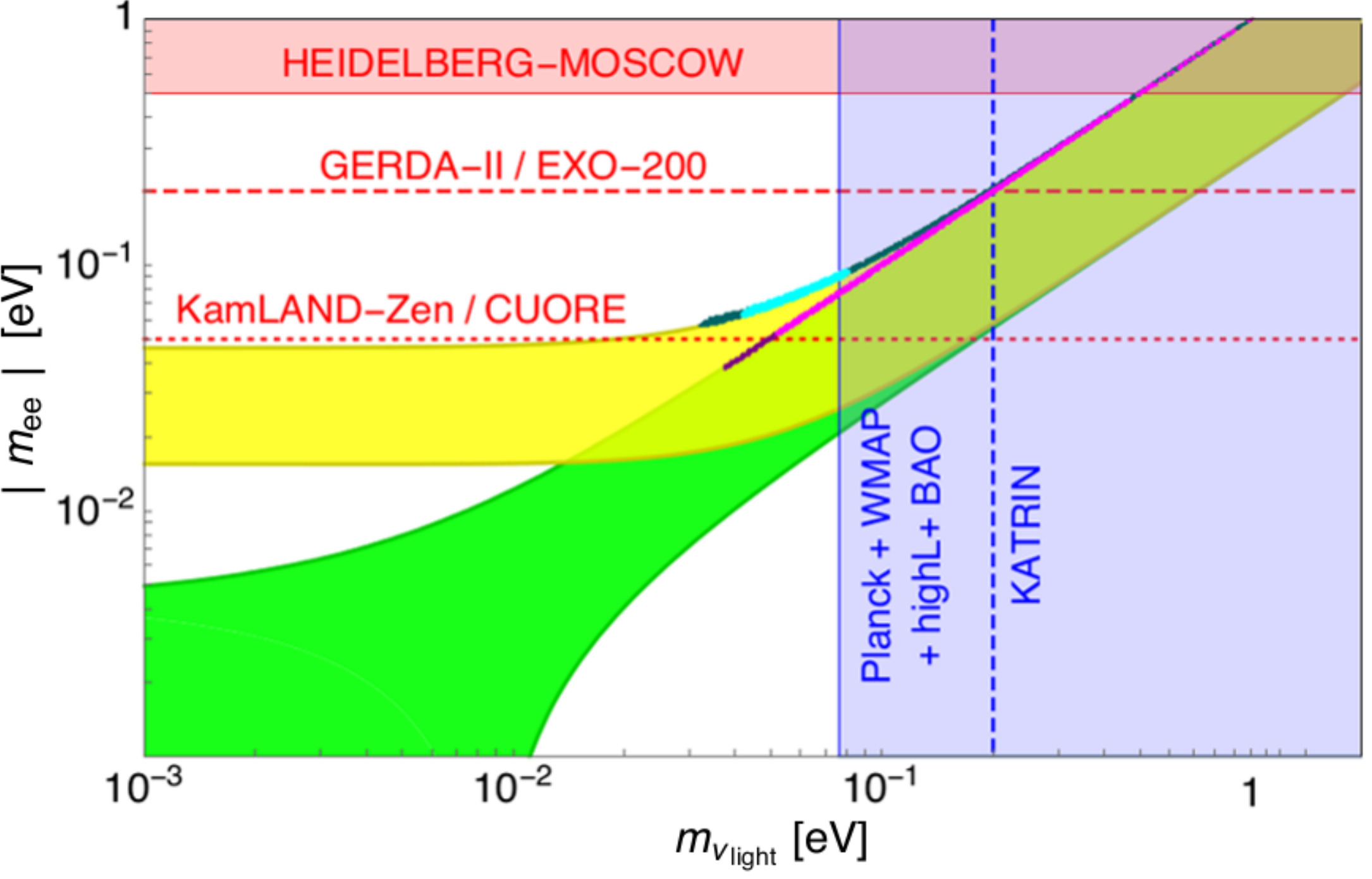} \includegraphics[scale=0.37]{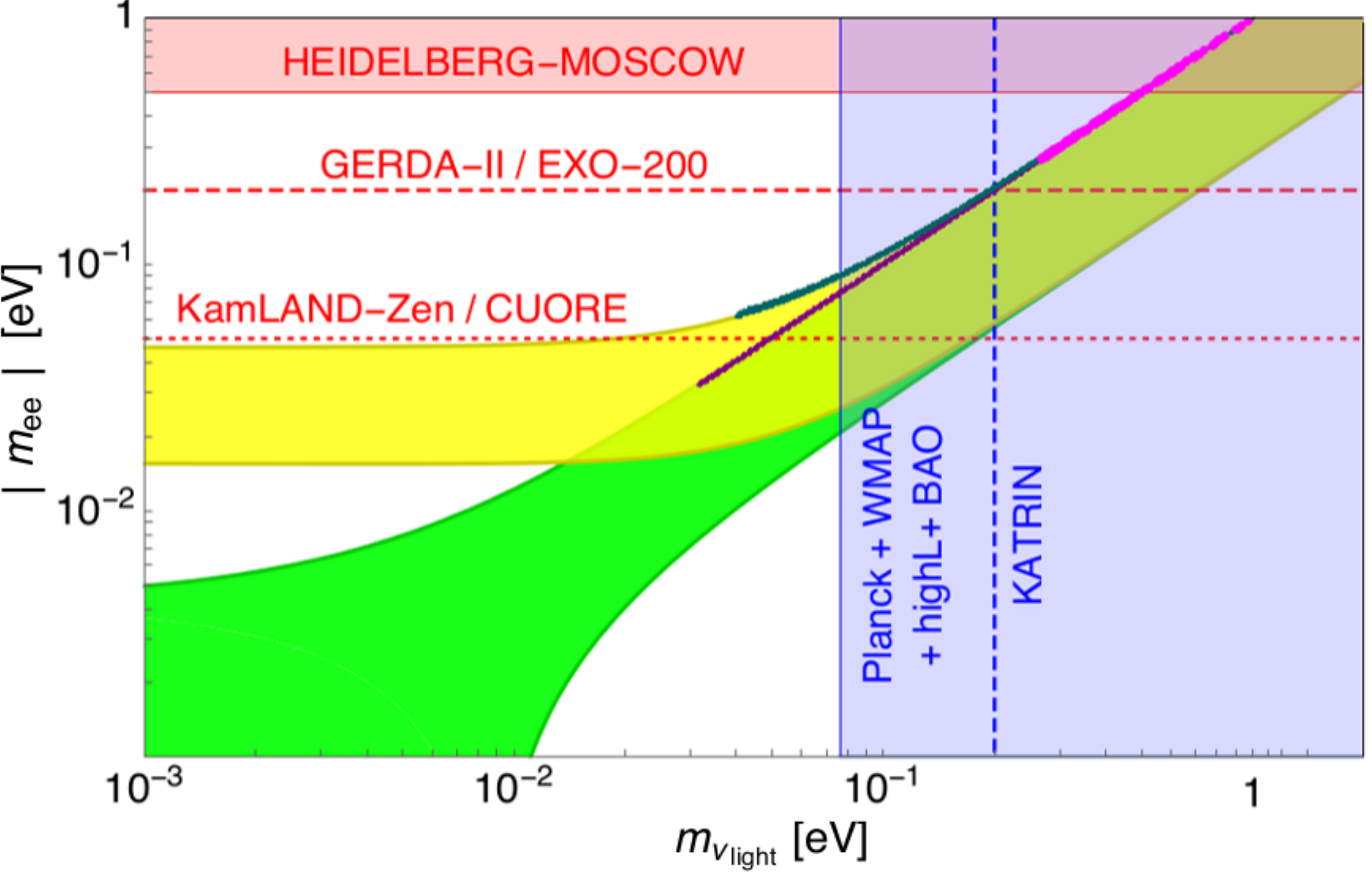}
\caption{Effective $0\nu\beta\beta$ parameter $|m_{ee}|$ versus the lightest neutrino mass $m_{\nu_\text{light}}$ in model A (B) on the left (right). In both models $m_{\nu_\text{light}}$ is $m_1$($m_3$) for NH (IH). The model allowed regions for NH is in magenta (dark magenta) for the $1\sigma$ ($3\sigma$) for the atmospheric mixing angle and for IH in cyan (dark cyan) for the $1\sigma$ ($3\sigma$) for the atmospheric mixing angle region from~\cite{Gonzalez-Garcia:2015qrr}. The horizontal blue (red) shaded region is the $1\sigma$ in $\sin^2 \theta_{23}$ for IH (NH). The yellow (green) band correspond to the $3\sigma$ ``flavor-generic" inverse (normal) hierarchy neutrino spectra. The horizontal red shaded region is the current experimental limit on $0\nu\beta\beta$, and the red (blue) horizontal (vertical) lines are the forthcoming experimental sensitivity on $|m_{ee}|$ ($m_{\nu_\text{light}}$), see~\cite{Agostini:2013mzu,Albert:2014awa,Gando:2012zm,Bornschein:2003xi,CUORE,Guiseppe:2008aa}.  The vertical blue shaded regions are disfavored by the current Planck data~\cite{Ade:2015xua}.}
\label{fig:graf3}
\end{figure}

\begin{figure}[!h]
\centering
\includegraphics[scale=0.37]{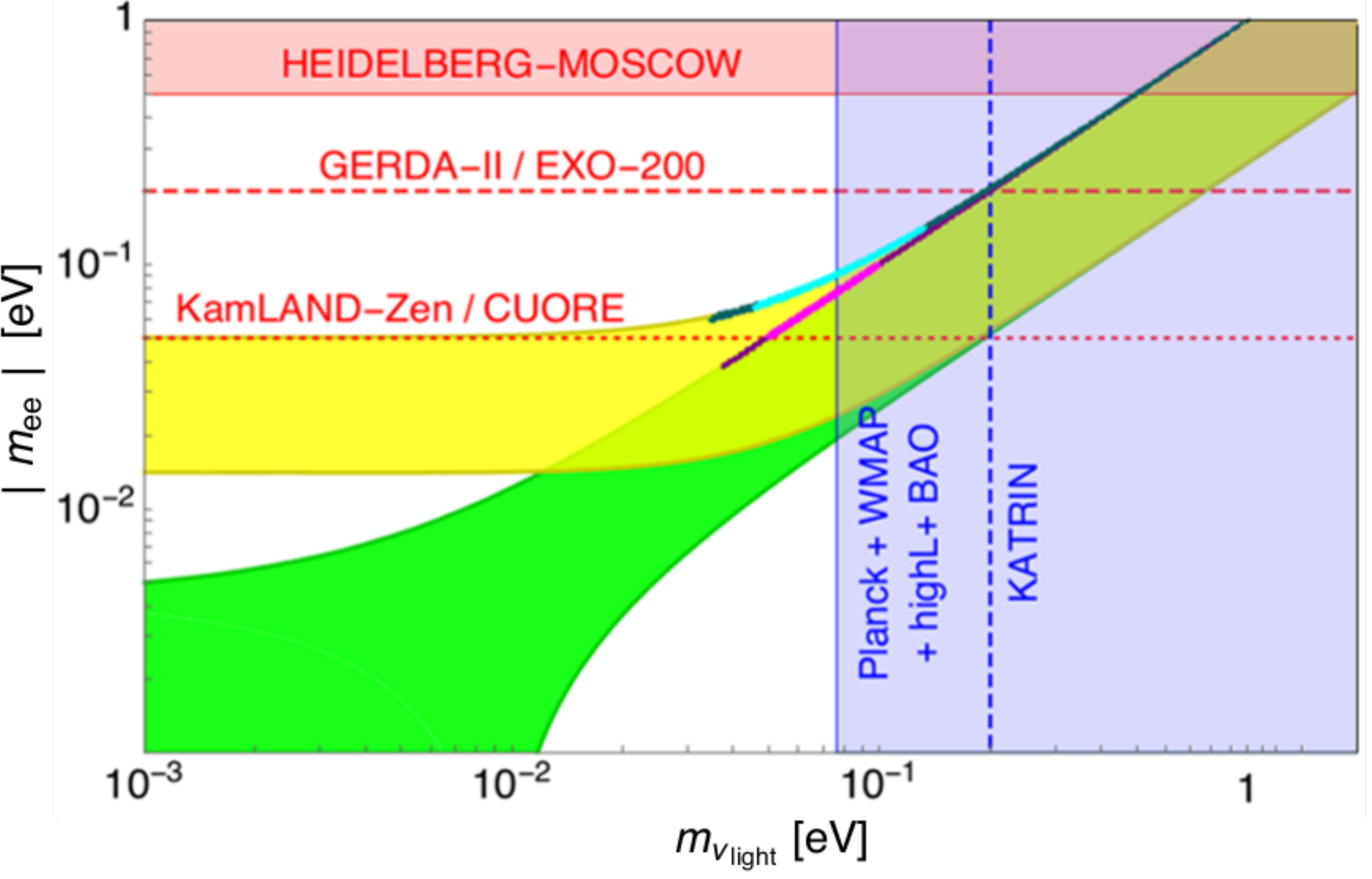} \includegraphics[scale=0.37]{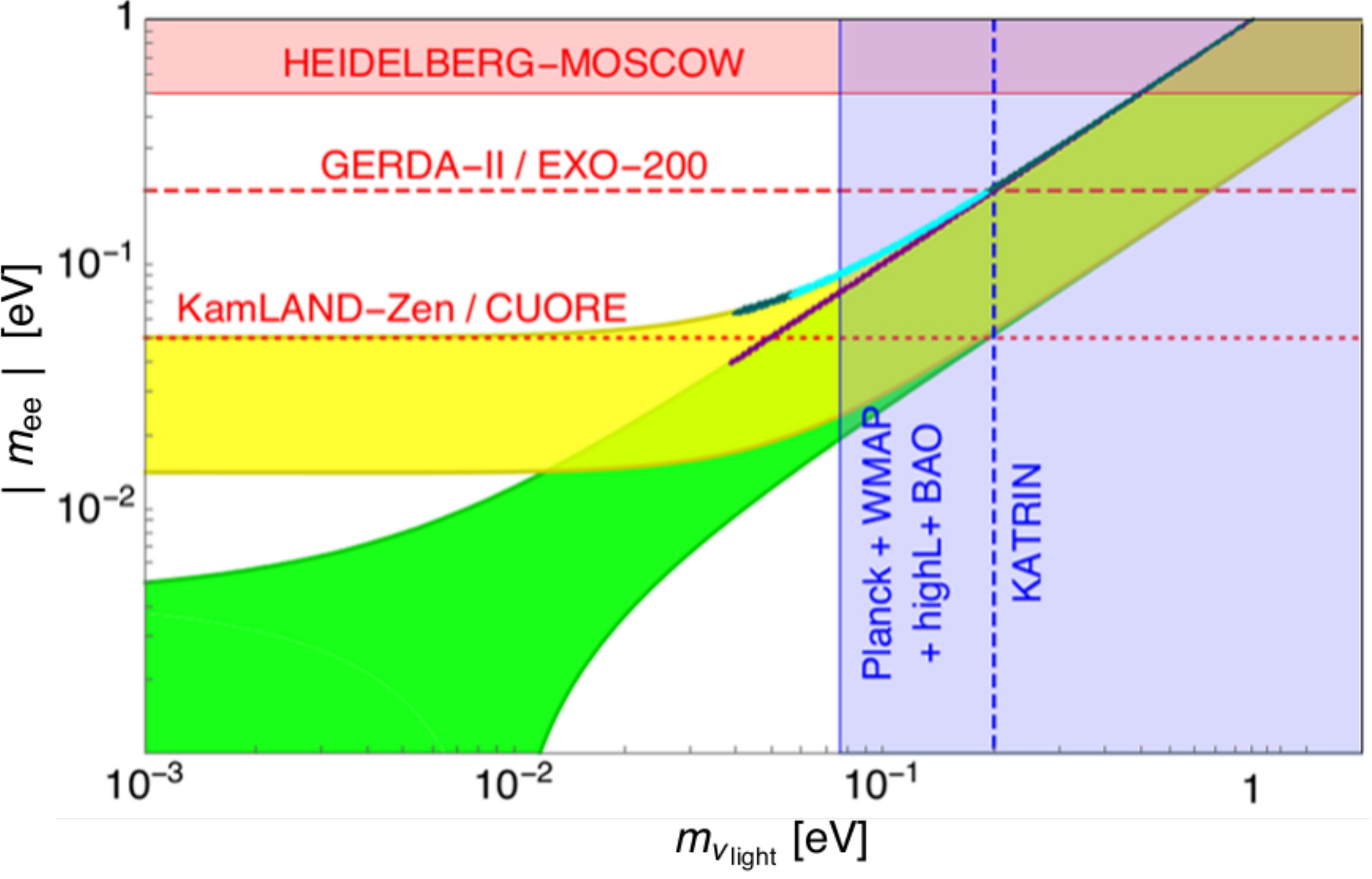}
\caption{Effective $0\nu\beta\beta$ parameter $|m_{ee}|$ versus the lightest neutrino mass $m_{\nu_\text{light}}$ in model A (B) on the left (right). Regions for NH is in magenta (dark magenta) for the $1\sigma$ ($3\sigma$) atmospheric mixing angle region and the IH in cyan (dark cyan) for the $1\sigma$ ($3\sigma$) for the atmospheric mixing angle region from~\cite{Capozzi:2016rtj}. The yellow (green) band correspond to the $3\sigma$ ``flavor-generic" inverse (normal) hierarchy neutrino spectra. The horizontal red shaded region is the current experimental limit on $0\nu\beta\beta$, and the red (blue) horizontal (vertical) lines are the forthcoming experimental sensitivity on $|m_{ee}|$ ($m_{\nu_\text{light}}$), see~\cite{Agostini:2013mzu,Albert:2014awa,Gando:2012zm,Bornschein:2003xi,CUORE,Guiseppe:2008aa}. The vertical blue shaded regions are disfavored by the current Planck data~\cite{Ade:2015xua}.}
\label{fig:graf5}
\end{figure}

The other correlation we obtained in the models  is  the neutrinoless double beta decay effective mass parameter, $|m_{ee}|$, with the lightest neutrino mass, $m_{\nu_\text{light}}$, where $m_{\nu_\text{light}} = m_{\nu_1}$ in the normal hierarchy and  $m_{\nu_\text{light}} = m_{\nu_3}$ in the inverted hierarchy. Figures~\ref{fig:graf1},~\ref{fig:graf3}, and~\ref{fig:graf5} show $m_{\nu_\text{light}}$ versus $|m_{ee}|$  for model A ($B_3$) on the left panels and model B ($B_4$) on the right ones. The region for the NH within $3\sigma$ are in dark magenta and the overlap for the atmospheric mixing angle of  1$\sigma$ in magenta; similarly, the region corresponding  to the IH within $3\sigma$ are in dark cyan and within $1\sigma$ in cyan.  The horizontal red shaded region corresponds the current experimental limit on neutrinoless double beta decay~\cite{Guiseppe:2008aa}; the red and blue lines are the forthcoming experimental sensitivities on $|m_{ee}|$~\cite{Agostini:2013mzu,Albert:2014awa,Gando:2012zm,CUORE} and $m_{\nu_\text{light}}$~\cite{Bornschein:2003xi}, respectively. The vertical blue shaded region is disfavored by the current Planck data~\cite{Ade:2015xua}.  In the graphics, we also show in yellow and green the bands corresponding to the $3\sigma$ ``flavor-generic" inverse and normal hierarchy neutrino spectra, respectively.

It can be seen from Figs.~\ref{fig:graf1} and~\ref{fig:graf3} that for model B there is no $1\sigma$  overlap between the prediction and the experimental data for the atmospheric mixing angle and therefore, we only show the data for the $3\sigma$ regions in the IH. In Fig. \ref{fig:graf5}, it can be seen that also the results in model B do not overlap with the 1$\sigma$ region for the NH case, as we mentioned before.  The models predict Majorana phases giving a minimal cancelation for the $|m_{ee}|$, as can bee seen in Figs.~\ref{fig:graf1},~\ref{fig:graf3}, and~\ref{fig:graf5}. The allowed regions for the $|m_{ee}|$ are in the upper lines for NH and IH generic bands. The two-zero textures  $B_3$ and $B_4$ are sensitive to the value of the atmospheric mixing angle. In the cases in which the atmospheric mixing angle prediction overlaps with the experimental value at $1\sigma$,  it translates to a localized region for neutrinoless double beta decay  within the near future experimental sensitivity, which is a desirable feature. A better measurement of the atmospheric mixing angle would be crucial for this kind of scenarios.

The dark matter phenomenology arising from the models (A and B) is different from that in the original DDM model, where the limit for large masses ($M_\text{DM} > 100$ GeV) was not allowed. The DM phenomenology is similar to the one in the inert Higgs doublet model~\cite{Deshpande:1977rw} with  two active and two inert Higgses.  What can be said about the DM phenomenology is that there is no inconvenient in generating the correct relic abundance even if the mass of the DM candidate is bigger than the mass of the gauge bosons. The limits presented in the minimal dark matter model~\cite{Cirelli:2005uq} apply, and for those masses, it annihilates mainly into gauge bosons. 

Finally, it is worth mentioning that neutrino phenomenology and the dark matter phenomenology are related by the way $A_4$ is broken into the $Z_2$ symmetry. This breaking dictates the pattern of masses and the mixing of the neutrinos, and at the same time, this $Z_2$ is responsible for the DM stability. This is the connection between DM and neutrinos in the presented models. 

\section{Conclusions}
\label{sec:Con}

We have constructed two models based on the discrete dark matter mechanism  where the non-Abelian $A_4$ flavor symmetry is spontaneously broken at the seesaw scale, into a remanent $Z_2$. In these models, we have  a total of five RH neutrinos. In this case, two RH neutrinos are in the $Z_2$ odd sector, and  the other three RH are even under  $Z_2$. These three RH neutrinos are responsible for giving the light neutrino masses via type I seesaw. Additionally, we have added  flavon scalar fields $\phi$ leading to the $A_4$ breaking in such a way that we obtained two-zero textures for the light Majorana neutrinos. These textures give rise to rich neutrino phenomenology: the results are in agreement  with the experimental data of the reactor mixing angle and accommodate the two possible neutrino mass hierarchies, NH and IH. 

Another consequence of the way $A_4$ is broken, in addition to dictating the neutrino phenomenology, is that these models contain a DM candidate stabilized by the remnant $Z_2$ symmetry. The DM phenomenology in this case will be different than the original DDM \cite{Boucenna:2011tj}, where the limit for large DM masses ($M_{DM}\gtrsim 100$ GeV) was not allowed, and will be similar to the inert Higgs doublet model~\cite{Deshpande:1977rw} with extra scalar fields.  A detailed discussion of the DM phenomenology is beyond the scope of the present work and will be presented in a further work~\cite{inprep}.
 
Additionally, we have updated the analysis for the two-zero textures mass matrix obtained for both models $B_3$ and $B_4$.  We presented the correlation between the atmospheric mixing angle and the sum of the light neutrino masses as well as the lower bounds for neutrinoless double beta decay effective mass parameter; the latter being in the region of sensitivity of the near future experiments. Finally, if the flavon fields acquire  vevs at a scale slightly higher than the seesaw scale, the remaining symmetry at the seesaw scale is the $Z_2$, and this would imply a mixing of the three $Z_2$ even RH neutrinos, which could be crucial if we want to have a scenario for leptogenesis, since in the $A_4$ symmetric case this was not possible.\footnote{This is studied somewhere else~\cite{inprep}}

\section*{ACKNOWLEDGEMENTS}

This work has been supported in part by Grants No. PAPIIT IA101516, No. PAPIIT IN111115, No. CONACYT 132059 and SNI. J.M.L.  would like to thank CONACYT (M\'exico) for financial support.
  
\appendix

\section{THE $A_4$ PRODUCT REPRESENTATION}
\label{a4rep}

The group $A_4$ has four irreducible representations: three singlets $\bf{1}$, $\bf{1^{\prime}}$, and $\bf{1^{\prime\prime}}$ and one triplet $3$ and two generators: $S$ and $T$ following the relations $S^2 = T^3 = (ST)^3 = \mathcal{I}$.  The one-dimensional unitary representations are
\begin{equation}
\begin{array}{lll}
\bf{1}: & S=1, & T=1,\\
\bf{1^{\prime}}: &S=1,&T=\omega,\\
\bf{1^{\prime\prime}}:&S=1,&T=\omega^2,
\end{array}
\end{equation}
where $\omega^3=1$.
In the basis where $S$ is real diagonal,
\begin{equation}
\label{eq:ST}
S=\left(
\begin{array}{ccc}
1&0&0\\
0&-1&0\\
0&0&-1\\
\end{array}
\right)\,\text{and}\quad
T=\left(
\begin{array}{ccc}
0&1&0\\
0&0&1\\
1&0&0\\
\end{array}
\right)\,. 
\end{equation}

The product rule for the singlets are
\begin{equation}
\begin{array}{l}
{\bf 1}\times{\bf1}=\bf{1^{\prime}}\times\bf{1^{\prime\prime}}=1,\\
\bf{1^{\prime}}\times\bf{1^{\prime}}=\bf{1^{\prime\prime}},\\
\bf{1^{\prime\prime}}\times\bf{1^{\prime\prime}}=\bf{1^{\prime}},
\end{array}
\end{equation} 
and triplet multiplication rules are
\begin{equation}\label{pr}
\begin{array}{lll}
(ab)_1&=&a_1b_1+a_2b_2+a_3b_3\,,\\
(ab)_{1'}&=&a_1b_1+\omega a_2b_2+\omega^2a_3b_3\,,\\
(ab)_{1''}&=&a_1b_1+\omega^2 a_2b_2+\omega a_3b_3\,,\\
(ab)_{3_1}&=&(a_2b_3,a_3b_1,a_1b_2)\,,\\
(ab)_{3_2}&=&(a_3b_2,a_1b_3,a_2b_1)\,,
\end{array}
\end{equation}
where $a=(a_1,a_2,a_3)$ and $b=(b_1,b_2,b_3)$.

\end{document}